%% file: squeezed-slit.tex
\documentclass[amsmath,amssymb,aps,prl,preprint]{revtex4-2}

	\usepackage{graphicx}
	\usepackage{soul}
	\usepackage[colorlinks=true,citecolor=blue,linkcolor=magenta]{hyperref}
	\usepackage[usenames]{color}
	\usepackage{amsfonts}
	\usepackage{booktabs}
	\usepackage{multirow}
	\usepackage{float}
    \usepackage[english]{babel}
	\usepackage{textcomp}
	\usepackage{upgreek}
    \usepackage{physics}
    \usepackage{siunitx}

\begin{document}

\title{Squeezed-slit Bohr-Einstein Interferometer}

\author{Hao-Wen Cheng$^{1,2,3}$, 
Xu-Zhao-Qiu Zeng$^{1,2,3}$, 
Yu-Chen Zhang$^{1,2,3}$,
Yu-Hao Deng$^{1,2,3}$,
Zhan Wu$^{1,2,3}$,
Rui Lin$^{1,2,3}$,
Yu-Cheng Duan$^{1,2,3}$,
Zi-Han Chen$^{1,2,3}$,
Jun Rui$^{1,2,3}$,
Ming-Cheng Chen$^{1,2,3}$,
Chao-Yang Lu$^{1,2,3}$,
Jian-Wei Pan$^{1,2,3}$}

\affiliation{$^1$Hefei National Laboratory for Physical Science at the Microscale and Department of Modern Physics, University of Science and Technology of China, Hefei 230026, China}
\affiliation{$^2$Shanghai Branch, CAS Center for Excellence in Quantum Information and Quantum Physics, University of Science and Technology of China, Shanghai 201315, China}
\affiliation{$^3$New Cornerstone Science Laboratory, Shanghai Research Center for Quantum Sciences, Shanghai 201315, China}

\begin{abstract}
The Einstein-Bohr recoiling-slit gedankenexperiment, a cornerstone of quantum complementarity, has long been constrained by the zero-point fluctuations of the atomic slit---the spatial Standard Quantum Limit (SQL). Here we transcend this fundamental boundary through active quantum state engineering of a single-atom slit. By implementing a non-adiabatic quench-evolve-quench protocol, we prepare the atomic motion in a squeezed state, dynamically redistributing phase-space uncertainty to suppress which-path information and restore high-visibility interference beyond the static vacuum limit. We report an intrinsic visibility of $0.938_{-0.008}^{+0.004}$, violating the SQL ($0.819$) by over 10 standard deviations, corresponding to $7.6(2)$~\si{dB} of effective squeezing. Our work reveals Kerr-induced non-Gaussian dynamics and reinterprets the traditional interferometer as a powerful tool for continuous-variable Wigner tomography, bridging the gap between quantum foundations and advanced metrology.
\end{abstract}

\maketitle

\begin{figure*}[ht!]
	\centering
	\includegraphics[width=\textwidth]{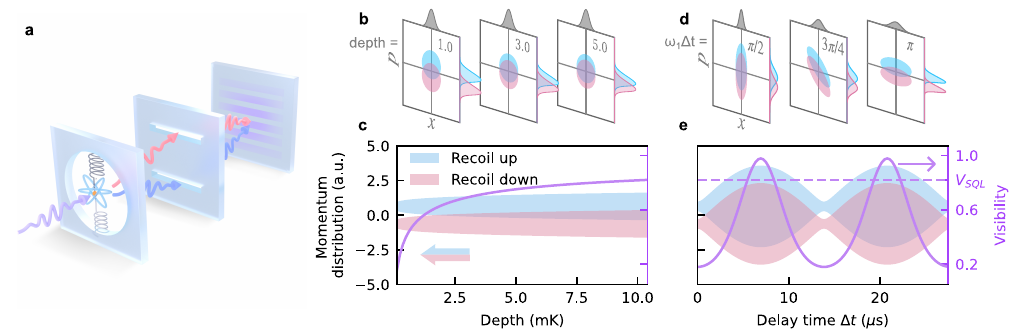}
    	\caption{\textbf{Conceptual illustration of the Einstein-Bohr gedankenexperiment with a single trapped atom.} 
    	\textbf{(a)} The atom serves as a movable slit. Photon scattering imparts opposite recoil momenta $\pm\hbar k$, entangling the optical path with the atomic motion. 
        \textbf{(b)} Phase-space representation of the atomic slit in the isotropic ground state of a static potential. The Wigner function (center) and its marginal distributions in position ($x$, top) and momentum ($p$, right) are shown. Blue and red regions denote the motional states correlated with the two photon paths. From left to right, the trap depth increases (1.0, 3.0, 5.0~\si{mK}), reducing the relative recoil displacement in normalized phase space.
        \textbf{(c)} Interference visibility in the passive static regime, determined by the overlap between the two recoil-induced motional wave packets (blue and red) in momentum space. The visibility (purple curve) increases with trap depth but remains bounded by the SQL set by ground-state fluctuations.
        \textbf{(d)} Phase-space representation of a motionally squeezed state in a fixed potential undergoing dynamical rotation. From left to right, the panels show increasing dynamical rotation angle $\omega_1 \Delta t$ ($\pi/2$, $3\pi/4$, $\pi$), where $\omega_1$ is the trap frequency governing the evolution.
        \textbf{(e)} Interference visibility in the active dynamic regime. Phase-space rotation of the squeezed state periodically maximizes the momentum uncertainty, allowing the visibility to transiently surpass the static SQL.
    }
	\label{fig1}
\end{figure*}

The 1927 Solvay Conference debate between Einstein and Bohr established complementarity as a central tenet of quantum mechanics. The essence of their recoiling-slit gedankenexperiment lies in the trade-off between path information and interference: when a photon scatters off a movable slit, it transfers recoil momentum, thereby marking its path and erasing the interference fringes \cite{bohr1949discussion,bacciagaluppi2009quantum,wheeler1983quantum,wootters1979}. While this foundational concept has been realized in various physical systems \cite{eichmann1993young,chapman1995,durr1998,eschner2001light,kokorowski2001single,bertet2001,chang2008quantum,tomkovivc2011single,schmidt2013momentum,weisz2014electronic,margalit2015self,liu2015einstein}, recent implementations using single trapped atoms have successfully mapped the transition from wave-like to particle-like behavior by tuning the static trapping potential \cite{zhang2024tunable} (Fig.~\ref{fig1}a). 

However, these passive approaches are fundamentally constrained by the zero-point fluctuations of the atom's motional ground state (Fig.~\ref{fig1}b). We define this minimum position uncertainty in a static potential as the spatial Standard Quantum Limit (SQL) of the atomic slit \cite{caves1981quantum, giovannetti2004quantum}. While the particle-like regime is easily accessible by relaxing the confinement, preserving the photon's wave-like character (i.e., achieving near-unity visibility) requires drastically compressing the atomic motional wavepacket (Fig.~\ref{fig1}c). Surpassing the SQL would theoretically necessitate an infinitely deep potential, which is experimentally prohibited by finite laser power and recoil-induced decoherence \cite{turchette2000decoherence, leibfried2003quantum}.

Here, we transcend the limitation of static confinement through active quantum state engineering \cite{lienhard2025generation, leibfried2003quantum, meekhof1996generation}. Instead of passively tuning trap parameters, we implement a non-adiabatic quench-evolve-quench protocol \cite{xin2021rapid} to prepare the atomic slit in a motionally squeezed state. This dynamic operation redistributes phase-space uncertainty. Reducing the spatial uncertainty below the ground-state limit ($\Delta x_\text{SQL}$) leads to a corresponding increase in the momentum spread (Fig.~\ref{fig1}d). Such momentum anti-squeezing obscures the photon recoil, erasing the which-path information and restoring high-visibility interference beyond the static SQL \cite{scully1991quantum} (Fig. \ref{fig1}e).

A Bayesian reconstruction yields a peak intrinsic visibility of $0.938^{+0.004}_{-0.008}$ at zero temperature \cite{foreman2013emcee}, surpassing the vacuum-limit visibility bound ($V_{\text{SQL},0} = 0.819$) set by the ground-state fluctuations of the atomic slit by over 10 standard deviations. From the reconstructed intrinsic dynamics, we infer an effective squeezing of $7.6(2)$~\si{dB}. Crucially, the interference visibility equals the modulus of the quantum characteristic function of the atomic motional state, evaluated at a fixed phase-space displacement \cite{lipkin1960some, weedbrook2012, sm}. This equivalence allows the interferometer to directly access the characteristic function, providing a route to continuous-variable Wigner tomography \cite{fluhmann2020}. During extended evolution, the visibility envelope exhibits a gradual decay arising from the intrinsic anharmonicity of the optical tweezer. We capture this behavior using a phenomenological Kerr Hamiltonian \cite{milburn1986quantum}, which accounts for the resulting phase-space shearing and enables extraction of the intrinsic dynamics.

\begin{figure*}[ht!]
	\centering
	\includegraphics[width=\textwidth]{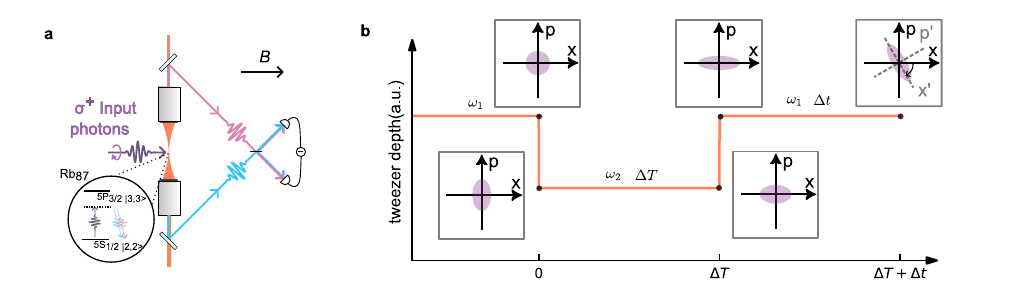}
	\caption{\textbf{Experimental setup and squeezed-state generation protocol.} 
	\textbf{(a)} A single $^{87}\text{Rb}$ atom is trapped in an optical tweezer. Confocal high-NA objectives collect photons scattered into opposing axial directions (red/blue paths), imparting opposite recoil momenta $\pm\hbar k$. \textit{Inset}: Energy-level diagram of the Rayleigh scattering transition ($5S_{1/2} \ket{2,2} \leftrightarrow 5P_{3/2} \ket{3,3}$).
    \textbf{(b)} Time sequence of the quench-evolve-quench (QEQ) protocol. At $t=0$, the trap frequency is switched from the deep potential $\omega_1$ to the shallow potential $\omega_2$ (a quench down). After an evolution time $\Delta T$ corresponding to a quarter-period rotation, the initial position squeezing is converted into momentum squeezing (bottom insets). The trap depth is then abruptly restored to $\omega_1$ (quench up). This second quench projects the expanded wavepacket onto the stiff potential, effectively amplifying the squeezing. \textit{Upper-right inset}: Phase-space rotation during the subsequent evolution time $\Delta t$ in the deep trap.
	}
    \label{fig2}
\end{figure*}

\textit{Experiment and Results---}Our experiment employs a single $^{87}\text{Rb}$ atom trapped in an 852-nm optical tweezer (NA=0.55) \cite{zhang2024tunable}. Raman sideband cooling prepares the atom in the three-dimensional motional ground state \cite{kaufman2012, thompson2013, yu2018motional, zhang2024tunable}. A radial probe beam drives Rayleigh scattering, imparting a momentum recoil of $\pm \hbar k$ along the longitudinal axis (Fig.~\ref{fig2}a). This process applies the momentum-displacement operator $e^{\mp i k\hat{x}}$, entangling the atomic motion with the photon path:
\begin{equation*}
    \ket{\Psi_\text{final}}=\frac{1}{\sqrt{2}}\left(e^{-ik\hat{x}}\ket{\psi_i}\otimes\ket{\text{ path}_1}+e^{+ik\hat{x}}\ket{\psi_i}\otimes\ket{\text{path}_2}\right),
\end{equation*}
where $\ket{\psi_i}$ denotes the initial motional state.

The interference visibility $V$ quantifies the overlap between the two recoil-shifted motional wavepackets. For an arbitrary motional state $\rho$, the visibility equals the modulus of the characteristic function evaluated at the recoil separation $\xi=2k$ \cite{sm}:
\begin{equation*}
    V=\abs{\tr(\rho e^{2ik\hat{x}})}=\abs{\chi(2k)},
\end{equation*}
where $\chi(\xi)=\tr(\rho e^{i\xi \hat{x}})$ \cite{lipkin1960some,weedbrook2012}. Static confinement limits this overlap. For an atom in the ground state of our deep trap ($\omega_1 = 2\pi\times 37.8(3)$~\si{kHz}), the zero-point spatial spread $\Delta x_\text{SQL}$ sets a strict zero-temperature visibility bound: $V_{\text{SQL},0}=\exp(-2\eta^2)=0.819(1)$ \cite{sm}, where $\eta\equiv k\Delta x_\text{SQL}$ is the Lamb-Dicke parameter \cite{leibfried2003quantum}. Under realistic finite-temperature conditions, the initial thermal occupation further reduces the coherence, yielding the thermal limit $V_{\text{SQL},T}=\exp[-2\eta^2(2\bar{n}+1)]$ \cite{sm}, where $\bar{n}$ denotes the mean phonon number.

To surpass these static bounds, we employ a non-adiabatic quench-evolve-quench (QEQ) protocol to dynamically reshape the motional state (Fig.~\ref{fig2}b) \cite{janszky1986squeezing, xin2021rapid,sutherland2021motional,lienhard2025generation}. Reducing the position uncertainty along the recoil axis to $\Delta x_\text{SQ}=\Delta x_\text{SQL} e^{-S}$, where $S$ denotes the squeezing parameter, increases the visibility to $V(S)=\exp(-2\eta^2 e^{-2S})$ \cite{sm}. The sequence begins in the deep trap ($\omega_1$). A rapid quench to a shallow potential ($\omega_2 \approx 2\pi\times 12.8$~\si{kHz}) generates an initial squeezing parameter $S_1=\ln\sqrt{\omega_1/\omega_2}\approx0.54$. After a quarter-period evolution ($\Delta T=\pi/(2\omega_2)$), the phase-space distribution rotates by $\pi/2$, converting the initial position squeezing into momentum squeezing \cite{xin2021rapid, brandstetter2025magnifying, asteria2021quantum}. A second quench restores the trap frequency to $\omega_1$, projects the expanded wavepacket back onto the stiff potential, and effectively doubles the squeezing parameter to $S_2=2S_1\approx 1.08$ \cite{xin2021rapid}. Subsequent harmonic evolution continuously rotates the distribution, periodically producing the spatial compression required to suppress which-path information.

\begin{figure*}[t!]
	\centering
	\includegraphics[width=\textwidth]{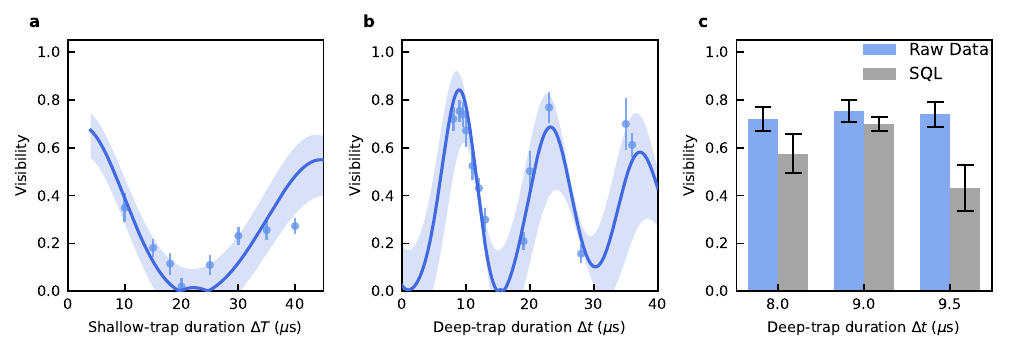}
	\caption{\textbf{Observed squeezing dynamics and enhancement beyond the thermal visibility limit.} 
	\textbf{(a)} Visibility as a function of the shallow-trap duration $\Delta T$ (with fixed deep-trap evolution time $\Delta t$). The visibility dip corresponds to the state of maximum spatial expansion and momentum squeezing.
    \textbf{(b)} Visibility as a function of the deep-trap duration $\Delta t$ (with fixed $\Delta T$). The oscillations reflect the phase-space rotation of the squeezed state, while the decay of the envelope indicates anharmonic phase-space shearing. The deep minima observed in both (a) and (b) approach zero and are consistent with Kerr-induced non-Gaussian dynamics.
    In panels (a) and (b), blue circles denote the measured raw visibility. The solid curves show the theoretical model including Kerr nonlinearity for a representative thermal occupation ($\bar{n}=0.5$), used as a visual guide. The shaded regions indicate prediction bands corresponding to the range of experimentally measured thermal occupations ($\bar{n}_\text{min}$ to $\bar{n}_\text{max}$) determined independently for each dataset. The simulations use parameters obtained from the global Bayesian analysis (single-quench squeezing parameter $S_1=0.50$, Kerr coefficient $K/\omega_1=-0.011$).
    \textbf{(c)} Benchmark against the theoretical thermal visibility limit. Blue bars show the measured raw visibility for the first peak in (b), while gray bars indicate the theoretical thermal limit $V_{\text{SQL},T}=\exp[-2\eta^2(2\bar{n}+1)]$, where $\eta$ is the Lamb-Dicke parameter and $\bar{n}$ is the independently measured thermal occupation for each data point. The measured values consistently exceed the corresponding $V_{\text{SQL},T}$. Error bars denote one standard deviation.
    }
	\label{fig3}
\end{figure*}

Experimental validation of this active protocol is revealed through the interference visibility. Mapping the visibility as a function of the shallow-trap duration $\Delta T$ directly characterizes the generation of squeezing (Fig.~\ref{fig3}a). With the subsequent deep-trap evolution $\Delta t$ fixed, the signal exhibits a pronounced dip near the quarter-period rotation ($\sim 19.5$~\si{\micro s}). This minimum corresponds to the moment of maximal spatial expansion and therefore maximal momentum squeezing of the state. Subsequent evolution in the restored deep trap produces high-contrast visibility oscillations (Fig.~\ref{fig3}b). These oscillations trace the phase-space rotation of the squeezed state, where visibility peaks occur when momentum anti-squeezing dominates, effectively burying the photon recoil and erasing the path information \cite{scully1991quantum, durr1998}.

\begin{figure*}[t!]
	\centering
	\includegraphics[width=\textwidth]{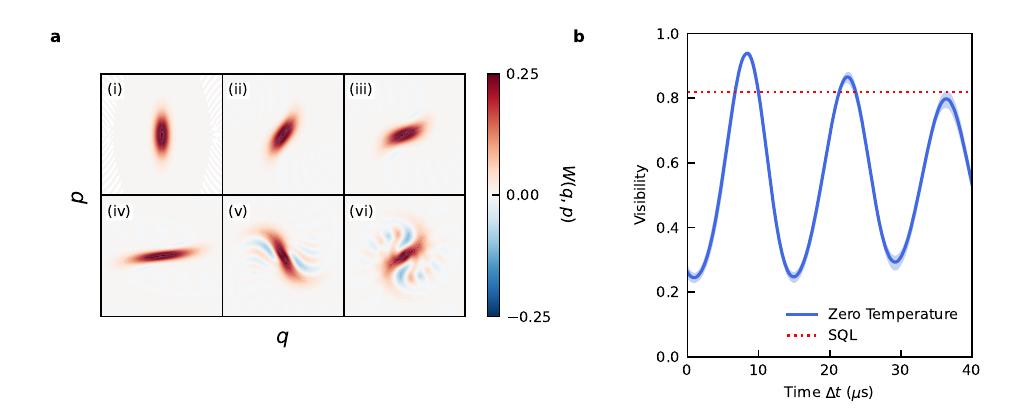}
	\caption{\textbf{Kerr-induced phase-space dynamics and intrinsic visibility beyond the SQL.} 
	Theoretical simulations based on the experimentally extracted parameters ($S_1=0.50(1)$ and $K/\omega_1=-0.011(2)$).
    \textbf{(a)} Evolution of the Wigner function $W(q,p)$ during the QEQ protocol.
    (i)--(iii) Dynamics in the shallow potential. 
    (i) The initial state is position-squeezed (vertically elongated) relative to the shallow potential.
    (ii),(iii) The state rotates in phase space under the anharmonic Hamiltonian.
    Panel (iii) shows the distribution near the optimal switching time $\Delta T$.
    (iv)--(vi) Dynamics in the restored deep potential.
    (iv) The state immediately after the second quench (quench up). Projection onto the stiff potential enhances the effective squeezing, producing a highly elongated distribution.
    (v),(vi) Continued evolution in the deep trap. Kerr nonlinearity induces phase-space shearing that progressively twists the distribution into an S-shaped spiral. The appearance of negative Wigner regions (blue) reflects the resulting non-Gaussian phase-space structure.
    \textbf{(b)} Reconstructed intrinsic zero-temperature visibility dynamics. The solid blue curve shows the simulated visibility evolution using the optimal parameters ($S_1$ and $K/\omega_1$), with the shaded region indicating the $1\sigma$ parameter uncertainty. The red dashed line marks the zero-temperature visibility bound $V_{\text{SQL},0}=0.819$. The reconstructed visibility periodically exceeds $V_{\text{SQL},0}$, while the decay of the oscillation envelope is consistent with coherent phase-space shearing. For comparison, the thermally corrected data obtained via power-law projection are shown in the Supplementary Material \cite{sm}.
    }
	\label{fig4}
\end{figure*}

Evaluating the first oscillation peak benchmarks this dynamical enhancement against the static limits. The raw interference visibility exceeds the thermal limit $V_{\text{SQL}, T}$ computed from independently measured mean phonon number (Fig.~\ref{fig3}c), confirming the active reshaping of the phase-space distribution. However, the initial finite temperature globally reduces the raw fringe contrast. To isolate the intrinsic quantum dynamics, we perform a global Bayesian analysis using the system Hamiltonian as a physical model \cite{foreman2013emcee, sm}. Marginalizing over the measured thermal occupation yields an intrinsic zero-temperature peak visibility of $V_\text{peak}=0.938_{-0.008}^{+0.004}$ (Fig.~\ref{fig4}b). This value exceeds the visibility bound $V_{\text{SQL},0}=0.819(1)$ by over 10 standard deviations, providing statistical confirmation of the quantum enhancement. Beyond surpassing this static bound, the active protocol expands the interferometer's operational range from a narrow passive window to a dynamic span of 0.22(2)--$0.938^{+0.004}_{-0.008}$, establishing the squeezed atomic slit as a tunable resource for exploring the full spectrum of wave-particle complementarity.

The subsequent evolution of this squeezed state reveals nonlinear dynamics of the optical tweezer. Throughout the observed phase-space rotations, the raw visibility envelope exhibits a gradual decay (Fig.~\ref{fig3}b). Given the negligible contributions from background heating and photon scattering ($\sim 1.7\%$ probability over \SI{40}{\micro s}) \cite{sm}, we attribute this damping primarily to the intrinsic anharmonicity of the trapping potential. We capture this behavior using an effective Kerr Hamiltonian $\hat{H}/\hbar\approx\omega_1\hat{n}+K\hat{n}^2$ \cite{milburn1986quantum, kitagawa1986number}, which produces an amplitude-dependent oscillation frequency $\omega(n)=\omega_1+2Kn$. The resulting phase-space shear distorts the distribution into a non-Gaussian geometry (Fig.~\ref{fig4}a). Consistent with this picture, the Kerr-induced distortion renders the spatial probability distribution non-Gaussian. Because the characteristic function is the Fourier transform of this distribution, the resulting $\chi(\xi)$ can develop oscillatory behavior in $\xi$-space (see Supplementary \cite{sm}). When evaluated at the interferometer probe point $\xi=2k$, such oscillations can lead to zero crossings of $\chi(2k)$, which manifest experimentally as visibility nodes in Fig.~\ref{fig3}. A global Bayesian analysis yields a Kerr coefficient $K/\omega_1=-0.011(2)$ together with a single-quench squeezing parameter $S_1=0.50(1)$ \cite{foreman2013emcee, sm}. These parameters reproduce the observed visibility dynamics within experimental uncertainty and are consistent with an interpretation based on Kerr-induced phase-space shearing \cite{sm}.

In summary, by transitioning the Einstein-Bohr recoiling-slit gedankenexperiment from passive tuning to active quantum state engineering, we achieved an intrinsic interference visibility of $0.938^{+0.004}_{-0.008}$, significantly surpassing the static SQL. This breakthrough, coupled with our analysis of Kerr-induced non-Gaussian dynamics, reinterprets the traditional interferometer as a powerful tool for continuous-variable Wigner tomography \cite{sm}. Beyond its foundational implications for quantum complementarity, this platform establishes a robust route for preparing and characterizing the non-Gaussian resources essential for advanced quantum information processing \cite{kendell2024deterministic, grochowski2025quantum,bohnmann2025bosonic}.

\begin{acknowledgments}
This work was supported by the National Natural Science Foundation of China (12322415), the Innovation Program for Quantum Science and Technology (2021ZD0301405), the HFNL Self-Deployed Project, and the New Cornerstone Science Foundation. The experiments utilized the ARTIQ control system \cite{bourdeauducq2021artiq}. H.-W.C., X.-Z.-Q.Z., and Y.-C.Z. contributed equally to this work.
\end{acknowledgments}

\bibliographystyle{apsrev4-2}
\bibliography{ref_num.bib}

\clearpage

\setcounter{equation}{0}
\setcounter{figure}{0}
\setcounter{table}{0}
\setcounter{page}{1}
\renewcommand{\theequation}{S\arabic{equation}}
\renewcommand{\thefigure}{S\arabic{figure}}
\renewcommand{\thetable}{S\Roman{table}}
\renewcommand{\thesection}{\Roman{section}}

\begin{center}
{\large\bf Supplemental Material for: ``Squeezed-slit Bohr-Einstein Interferometer''}
\end{center}

\setcounter{secnumdepth}{3}

\section{Experimental Implementation and Quench Dynamics}
\input{sup_section/experimental}

\section{Theoretical Calculation and Modeling}
This section will provide supplementary theoretical derivations.

\input{sup_section/theory_sec1v2}
\input{sup_section/theory_sec2v2}
\input{sup_section/theory_sec3v2}

\subsection{Effect of Kerr Anharmonicity on Visibility Calculation}
\input{sup_section/theory_sec4v2}

\section{Data Analysis and Correction}
\label{sec:data_analysis}
\subsection{Visibility Data Processing and Correction}
\input{sup_section/correction_sec1}

\subsection{Global Bayesian Parameter Estimation}
\input{sup_section/correction_sec2}

\section{Protocol for Interferometric Quantum State Tomography}
\input{sup_section/tomography}

\end{document}

%% file: sup_section/experimental.tex
The experimental apparatus and fundamental control sequences employed in this work are identical to those detailed in our previous work \cite{zhang2024tunable}. Specifically, we utilize the same single-atom optical tweezer setup (\SI{852}{nm}, NA=0.55), Raman sideband cooling (RSC) protocol for 3D ground-state initialization, and 1064 nm phase-locked Mach-Zehnder interferometer for stabilizing the optical path length. Furthermore, the real-time phase drift compensation scheme, which alternates between high-depth ground state reference cycles and science cycles to correct for slow thermal drifts using a ``moving window" technique, follows the exact procedure described previously. The thermal occupation $\bar{n}$ corresponding to each data is directly extracted from the Raman sideband spectrum.

Crucially, for the interference visibility measurement, we employ a fast detection window of \SI{1}{\micro s}. As characterized in Ref.~\cite{zhang2024tunable}, this timescale is sufficiently short to temporally resolve the fast motional oscillations and suppress recoil-induced decoherence during the measurement itself, ensuring a faithful readout of the instantaneous motional state.

To realize the squeezed state, we implement a non-adiabatic ``quench-evolve-quench" (QEQ) protocol. This requires rapid modulation of the trap depth, which is achieved via an acousto-optic modulator (AOM) controlling the tweezer intensity. We characterize the intensity ramp time to be $\tau_\text{ramp}\approx200$~\si{ns}. This timescale is two orders of magnitude shorter than the quarter-period evolution in the shallow potential ($\Delta T \approx 19.5$~\si{\mu s}). Consequently, the trap modulation is effectively instantaneous relative to the squeezing dynamics, rigorously justifying the sudden quench approximation utilized in the theoretical model presented in the main text and derived in the following sections.

We assess the impact of environmental decoherence by calculating the photon scattering rate $\Gamma_\text{sc}$ induced by the trapping laser. For a trap depth of $U_0=k_\text{B}\times 10.5$~\si{mK} at $\lambda=852$~\si{nm}, we sum the contributions from both the $\text{D}_1$ (\SI{795}{nm}) and $\text{D}_2$ (\SI{780}{nm}) lines of $^{87}\text{Rb}$:
\begin{equation}
    \Gamma_\text{sc} \approx \frac{\Gamma}{\hbar} \left( \frac{ \Delta_\text{D1}^{-2} + 2\Delta_\text{D2}^{-2} }{ \Delta_\text{D1}^{-1} + 2\Delta_\text{D2}^{-1} } \right) U_0,
\end{equation}
where $\Gamma=2\pi\times 6$~\si{MHz} is the natural linewidth, and $\Delta_\text{D1,D2}$ are the respective detunings. This yields an effective scattering rate of $\Gamma_\text{sc}\approx 280$~\si{s^{-1}}. Integrated over the maximum experimental sequence duration ($\tau_\text{seq}\approx 60$~\si{\micro s}), the total scattering probability is limited to $P_\text{sc}\approx 1.7\%$, confirming that vacuum scattering is a negligible source of decoherence.

%% file: sup_section/theory_sec1v2.tex
\subsection{General Relation between Visibility and Characteristic Function}
\label{sec:visibility_arbitrary}

This section establishes the fundamental link between the experimental measured interference signal and the quantum characteristic function of the atom's motional state.

The experiment employs a single-atom Mach-Zehnder interferometer. The observable is the normalized asymmetry of photon counts between the two output ports, $A$ and $B$, modulated by a controllable relative phase $\phi_\text{int}$:
\begin{equation} 
    A(\phi_{\text{int}}) = \frac{N(A) - N(B)}{N(A) + N(B)} 
    \label{eq:AsymmetryDef} 
\end{equation}
The amplitude of this oscillation, $A(\phi_\text{int})$, defines the intrinsic interference visibility $V$.

Physically, the interference arises from the scattering event, which entangles the photon's path degree of freedom ($\ket{\text{path}_1}$, $\ket{\text{path}_2}$) with the atom's motional state via momentum conservation. Emission into path 1 (path 2) imparts a momentum kick $-k$ ($+k$), represented by the displacement operators $e^{-ik\hat{x}}$ and $e^{+ik\hat{x}}$, respectively. For an initial atomic density matrix $\rho_i$, the scattering and subsequent recombination on a 50/50 beam splitter map the coherence between these two momentum-displaced states onto the photon detection probabilities.

Tracing out the atomic degrees of freedom yields the explicit form of the asymmetry signal:
\begin{equation} 
A(\phi_{\text{int}}) = \Im\left(e^{i\phi_{\text{int}}}\tr\left(\rho_i e^{2ik\hat{x}}\right)\right) 
\end{equation}
We identify the complex expectation value $\tr(\rho e^{2ik\hat{x}})$ as the quantum characteristic function of the motional state, $\chi(\xi)=\expval{e^{i\xi\hat{x}}}$, evaluated at the recoil momentum transfer $\xi=2k$. Writing this characteristic function in polar form as $\chi(2k)=Ve^{i\alpha}$, experimentally observed signal becomes a sinusoid:
\begin{equation}
    A(\phi_{\text{int}}) = V\sin(\phi_{\text{int}}+\alpha)
\end{equation}
Thus, the interference visibility $V$ directly probes the magnitude of the characteristic function at the specific spatial frequency defined by the interferometer geometry:
\begin{equation}
    V=\abs{\chi(2k)}=\abs{\tr(\rho_i e^{2ik\hat{x}})}
    \label{eq:V_final}
\end{equation}
This relation holds true for any arbitrary motional state $\rho$.

For the subsequent analysis, we adopt dimensionless operators. We define the characteristic length of the harmonic trap ground state as $x_0 = \sqrt{\hbar/2m\omega}$ and the dimensionless position operator $\hat{q}=\hat{x}/x_0=\hat{a}+\hat{a}^\dagger$. The momentum kick is quantified by the Lamb-Dicke parameter $\eta = kx_0$. In this notation, the visibility becomes:
\begin{equation}
    V=\abs{\chi(2\eta)}=\abs{\tr(\rho e^{2i\eta\hat{q}})}
    \label{eq:V_final_dimensionless}
\end{equation}
This general expression forms the basis for the analysis in the following sections.

%% file: sup_section/theory_sec2v2.tex
\subsection{Visibility for General Gaussian States}
\label{sec:gaussian_general}

We begin by establishing an analytic framework based on the properties of Gaussian states. In the ideal harmonic limit, the primary motional states prepared in this experiment---the ground state, thermal state, and squeezed state---all belong to the family of Gaussian states. While the actual experimental evolution introduces non-Gaussian distortions due to trap anharmonicity (as detailed later in Sec.~\ref{sec:kerr_anharmornicity}), this Gaussian formalism provides the fundamental physical intuition and the baseline mathematical relationships linking the visibility directly to the phase-space geometry.

A general single-mode Gaussian state $\rho_\text{G}$ is entirely characterized by its first-moment displacement vector $d$ and its second-moment covariance matrix (CM) $\mathbf{\sigma}$. In the dimensionless phase-space coordinates defined in Sec.~\ref{sec:visibility_arbitrary} ($\hat{q}=\hat{a}+\hat{a}^\dagger$, $\hat{p}=-i(\hat{a}-\hat{a}^\dagger)$), these are defined as:
\begin{equation} 
\mathbf{d}=\begin{pmatrix} \expval{\hat{q}}\\ \expval{\hat{p}} \end{pmatrix}, \quad 
\bm{\sigma} = \begin{pmatrix} 
    \sigma_{11} & \sigma_{12} \\ 
    \sigma_{21} & \sigma_{22} 
    \end{pmatrix} 
\end{equation}
where the matrix elements are the symmetrized variances $\sigma_{ij} = \frac{1}{2}\expval{\{\Delta\xi_i, \Delta\xi_j\}}$ with $\xi=(\hat{q}, \hat{p})^\mathrm{T}$. Specifically, $\sigma_{11}=\expval{(\Delta\hat{q})^2}$ represents the position variance.

The characteristic function $\chi_\text{G}(\eta)$ for a general Gaussian state has a known analytic form. Evaluated at the specific probe vector $\mathbf{u}=(2\eta, 0)^\mathrm{T}$ dictated by our interferometer geometry, it reads:
\begin{equation} 
    \chi_\text{G}(2\eta, 0) = \exp(-\frac{1}{2} \mathbf{u}^T \bm{\sigma} \mathbf{u} + i \mathbf{u}^T \mathbf{d}) = \exp(-2\eta^2\sigma_{11}) e^{2i\eta\expval{\hat{q}}} 
\end{equation}
Substituting this into the general visibility relation derived in Eq.~\eqref{eq:V_final_dimensionless}, $V=\abs{\chi(2\eta)}$, the phase factor $e^{2i\eta\expval{\hat{q}}}$ is eliminated by the absolute value. We thus arrive at a universal expression linking the interference visibility directly to the state's position variance:
\begin{equation} 
    V_\text{G} = \exp(-2\eta^2 \sigma_{11}) 
    \label{eq:master_visibility} 
\end{equation}
This result provides the unifying framework for our analysis: any Gaussian state's visibility is determined solely by how ``squeezed" or ``expanded" its position variance $\sigma_{11}$ is relative to the probe scale $\eta$.

%% file: sup_section/theory_sec3v2.tex
\subsection{Visibility of Specific Motional States}
\label{sec:specific_states}

We now apply the general result in Eq.~\eqref{eq:master_visibility} to the specific states relevant to our experiment.

\subsubsection{Zero-Temperature SQL (Ground State)}
For an atom in the motional ground state $\ket{0}$, the position variance is determined by the vacuum fluctuations. This defines the spatial Standard Quantum Limit ($\Delta x_\text{SQL}$) of the atomic slit. In our dimensionless convention, the normalized variance is unity: 
\begin{equation} 
\sigma_{11}^{(\text{gs})} = \expval{(\hat{a}+\hat{a}^\dagger)^2}{0} - \expval{(\hat{a}+\hat{a}^\dagger)}{0}^2 = 1 
\end{equation}
Substituting $\sigma_{11}^{(\text{gs})}=1$ into Eq.~\eqref{eq:master_visibility} determines the upper bound on visibility, which we denoted as the zero-temperature limit:
\begin{equation} 
    V_{\text{SQL},0} = e^{-2\eta^2} 
\end{equation}
This establishes the benchmark $V_\text{SQL}$ cited in the main text. The Lamb-Dicke parameter $\eta$ relates to the base trap frequency $\omega_1$ via $\eta=k\sqrt{\hbar/(2m\omega_1)}$, where $m$ is the mass of a single Rubidium-87 atom. Based on the independently measured trap frequency $\omega_1/2\pi=37.8(3)$~\si{kHz}, we determine the Lamb-Dicke parameter to be $\eta=0.316(1)$. Propagating this uncertainty into the visibility limit yields the benchmark value $V_{\text{SQL},0}=0.819(1)$.

\subsubsection{Thermal State}
A thermal state with mean phonon occupation $\bar{n}$ is an isotropic Gaussian state. Its variance is broadened by the thermal factor $(2\bar{n}+1)$ relative to the ground state:
\begin{equation} 
    \sigma_{11}^{(\text{th})} = 2\bar{n} + 1 
\end{equation}
Substituting this into Eq.~\eqref{eq:master_visibility} immediately yields the thermal visibility limit:
\begin{equation}
    V_{\text{SQL},T} = e^{-2\eta^2 (2\bar{n}+1)}
    \label{eq:V_th_theory}
\end{equation}
This confirms that thermal fluctuations exponentially suppress the interference visibility, establishing the pragmatic reduced visibility baseline ($V_{\text{SQL},T}$) for experimental realizations.

\subsubsection{Squeezed Vacuum State}
An ideal squeezed vacuum state $\ket{\zeta}=\hat{S}(\zeta)\ket{0}$ is generated by the squeezing operator with parameter $\zeta=S e^{i\theta_0}$. Under the harmonic trap evolution, the squeezing axis rotates in phase space. The resulting time-dependent position variance is given by:
\begin{equation} 
    \sigma_{11}^{(\text{sq})}(t) = \cosh(2S) - \sinh(2S)\cos\phi_\text{sq}(t) 
\end{equation}
where the instantaneous squeezing phase $\phi_\text{sq}(t) = \theta_0 - 2\omega t$ incorporates both the initial preparation angle $\theta_0$ and the dynamical rotation. Inserting this variance into Eq.~\eqref{eq:master_visibility} gives the time-dependent visibility:
\begin{equation} 
    V_\text{sq}(S, t) = \exp[-2\eta^2(\cosh(2S) - \sinh(2S)\cos\phi_\text{sq}(t))]
    \label{eq:theory_squeezed}
\end{equation}
The visibility reaches its maximum when the squeezed quadrature aligns with the position axis (i.e., $\cos\phi_\text{sq}(t)=1$), yielding $V_\text{max}=\exp(-2\eta^2e^{-2S})$. Since $e^{-2S}<1$ for any $S>0$, this rigorously proves that squeezing allows the visibility to surpass the ground-state limit $V_\text{gs}$.

\subsubsection{Squeezed Thermal State}
In a realistic experimental scenario, the squeezing operation acts on an initial thermal state $\rho_\text{th}$ with mean occupation $\bar{n}$, rather than a pure vacuum state. The resulting state is a squeezed thermal state.

Crucially, the initial thermal noise is isotropic, meaning its covariance matrix is proportional to the identity matrix, $\sigma^{(\text{th})}=\frac{1}{2}(2\bar{n}+1)\mathbb{I}$. Since the squeezing operation corresponds to a linear symplectic transformation in phase space, it transforms the covariance matrix linearly. Consequently, the position variance of the squeezed thermal state, $\sigma_{11}^{(\text{st})}$, effectively inherits the geometry of the squeezed vacuum variance, $\sigma_{11}^{(\text{sq})}$, but is globally scaled by the thermal factor:
\begin{equation}
    \sigma_{11}^{(\text{st})} = (2\bar{n}+1) \sigma_{11}^{(\text{sq})}
\end{equation}
Substituting this scaled variance into Eq.~\eqref{eq:master_visibility} yields:
\begin{equation}
    V_{\text{st}} = \left[ \exp\left(-2\eta^2 \sigma_{11}^{(\text{sq})}\right) \right]^{2\bar{n}+1}
\end{equation}
From this, we obtain the fundamental relationship linking the visibility of a measurable squeezed thermal state to that of an ideal squeezed vacuum:
\begin{equation}
    V_{\text{st}} = (V_{\text{sq}})^{2\bar{n}+1}
    \label{eq:st_sq_relationship}
\end{equation}
This analytical result demonstrates that finite temperature acts as a deterministic scaling factor on the logarithmic visibility. It implies that the essential phase-space rotation dynamics of the squeezed state are preserved even at finite temperatures, with the visibility reduced strictly by the power-law relation derived above.

%% file: sup_section/theory_sec4v2.tex
\label{sec:kerr_anharmornicity}
While the Gaussian formalism establishes a baseline intuition, the optical tweezer is formed by a diffraction-limited, tightly-focused beam (NA=0.55), with the incident laser over-filling the objective pupil. This strong focusing inherently generates a non-parabolic trapping potential. Microscopically, due to the inversion symmetry of the dipole trap, the lowest-order correction to the harmonic approximation is the quartic term. 

We analyze the system in the dimensionless coordinates $(\hat{q},\hat{p})$ defined in \ref{sec:visibility_arbitrary}, which satisfy the commutation relation $[\hat{q}, \hat{p}]=2i$. For clarity in the dynamical analysis, we adopt natural units where $\hbar=1$, such that the Hamiltonian represents an energy in frequency units. The Hamiltonian governing the atomic motion is:
\begin{equation}
    \hat{H} = \frac{\omega_0}{4}(\hat{p}^2+\hat{q}^2)+\omega_0 \lambda\hat{q}^4
\end{equation}
where $\lambda$ is the dimensionless anharmonicity parameter. To calibrate this microscopic parameter against the standard Kerr nonlinearity parameter $K$ commonly used in effective quantum models ($H_\text{eff}=K\hat{n}^2$), we invoke the Rotating Wave Approximation (RWA). By expanding the position operator $\hat{q}=\hat{a}+\hat{a}^\dagger$ and retaining only the number-conserving terms, the expectation value yields $\expval{\hat{q}^4}{\hat{n}}=6\hat{n}^2+6\hat{n}+3$. Identifying the quadratic $\hat{n}^2$ dependence with the effective Kerr term, we derive the mapping $\omega_0 \lambda (6\hat{n}^2)= K \hat{n}^2$, which yields the relation
\begin{equation}
    \lambda = \frac{K}{6\omega_0}
\end{equation}
Consequently, the anharmonic potential in our dimensionless coordinates is $V_\text{anh}(\hat{q})=(K/6)\hat{q}^4$.

To describe the system dynamics, we analyze the exact time evolution of the Wigner function $W(q,p,t)$ generated by the Hamiltonian $H(q,p)=H_\text{ho}+V_\text{anh}(q)$. The dynamics are governed by the Moyal equation, which involves the sine of the Poisson bracket operator:
\begin{equation}
    \frac{\partial W}{\partial t} =
    2H \sin\left( \overleftarrow{\partial_q}\overrightarrow{\partial_p} - \overleftarrow{\partial_p}\overrightarrow{\partial_q} \right) W
\end{equation}
This can be expanded as:
\begin{equation}
    \frac{\partial W}{\partial t} = \{H, W\}_{\text{PB}} +
        2 \sum_{n=1}^{\infty} \frac{(-1)^n}{(2n+1)!} (\partial_q^{2n+1} H) (\partial_p^{2n+1} W)
\end{equation}
Here, the first term $\{H, W\}_{\text{PB}}=2(\partial_q H\partial_p W-\partial_p H \partial_q W)$ is the classical Poisson bracket scaled by the commutation factor.

For our specific quartic potential $V_\text{anh}(q)= \frac{K}{6} q^4$, the derivatives of order 5 and higher vanish, causing the infinite Moyal series to truncate exactly after the first quantum correction term ($n=1$). The equation of motion separates into three distinct physical contributions:
\begin{widetext}
    \begin{equation}
        \frac{\partial W}{\partial t} = 
        \underbrace{\omega_0 \left( q\frac{\partial W}{\partial p} - p\frac{\partial W}{\partial q} \right)}_{\text{Harmonic Rotation}} + 
        \underbrace{\frac{4K}{3} q^3 \frac{\partial W}{\partial p}}_{\text{Anharmonic Shearing}} -
        \underbrace{\frac{4K}{3} q \frac{\partial^3 W}{\partial p^3}}_{\text{Quantum Correction}}
        \label{eq:effective_moyal_equation}
    \end{equation}
\end{widetext}

The first term describes the standard phase-space rotation at frequency $\omega_0$. The second term ($n=0$) drives classical anharmonic dynamics, inducing an amplitude-dependent phase accumulation rate that manifests as geometric shearing. To confirm this term accurately captures the relevant physics, we note that treating the nonlinearity as a perturbation, the classical frequency shifts quadratically with the amplitude, $\Delta\omega(A)\propto K A^2$. This dependence is structurally identical to the quantum prediction $\Delta\omega(n)\propto Kn$ via the semi-classical correspondence $A^2 \sim n$, confirming that the shearing term correctly captures the characteristic frequency shift of the Kerr nonlinearity.

The third term ($n=1$) represents the fundamental quantum correction that prevents the breakdown of the uncertainty principle. Note that the coefficient $-4K/3$ arises from the third-order derivative of the quartic potential ($\partial_q^3 V = 4Kq$) combined with the Moyal expansion factor ($-\frac{1}{3!}\cdot 2 = -1/3$). Under the sole influence of the classical shearing term, the phase-space distribution is advected along classical trajectories, a process that inherently preserving the non-negativity of an initially positive Wigner function. In contrast, the third term acts as a source of quantum dispersion that generates negative regions in the Wigner function. As the classical flow stretches and twists the distribution into increasingly thin spiral filaments, this quantum correction intervenes when structures approach the sub-Planck scale.

For our experimental parameters, we initialize the system in a squeezed thermal state ($r\sim 1$), assuming a thermal occupation $\bar{n}\sim 1$ to account for worst-case cooling imperfections, and observe the evolution under weak nonlinearity ($\abs{K}/\omega_0\sim 0.01$) for a short duration ($t<5\pi/\omega_0$). Within this regime, and specifically for the visibility observable defined in Eq.~\eqref{eq:V_final_dimensionless}, the phase-space distribution undergoes classical geometric distortion but remains sufficiently smooth to render the third-order derivative negligible. Consequently, we neglect the quantum correction term. This reduction simplifies the dynamics to the evolution of classical trajectories sampled from the initial Wigner distribution, which corresponds to the truncated Wigner approximation (TWA).

\begin{figure*}[t]
    \includegraphics[width=\textwidth]{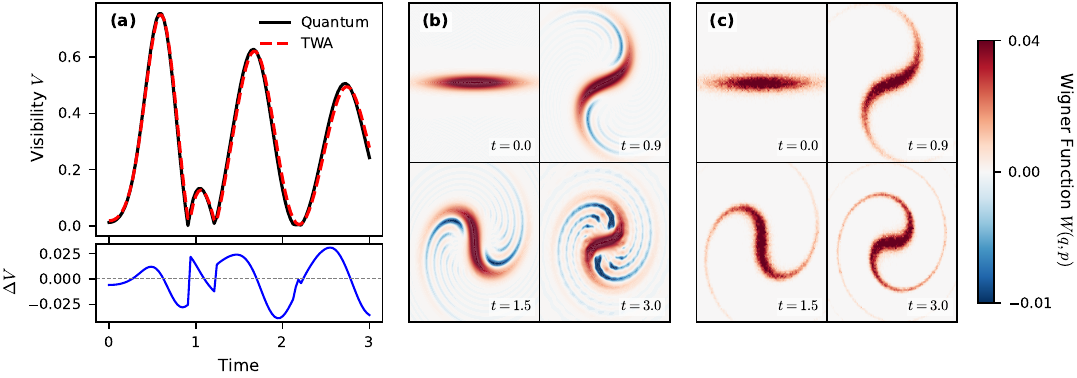}
    \caption{\textbf{Validation of Truncated Wigner Approximation (TWA).} 
    \textbf{(a)} Evolution of the interference visibility $V(t)=\abs{\expval{e^{2i\eta\hat{q}}}}$ with Lamb-Dicke parameter $\eta=0.316$. The system is initialized in a squeezed thermal state with mean occupation $\bar{n}=1.0$ and squeezing parameter $S=-1.0$ (anti-squeezed in $q$), evolving under a Kerr nonlinearity strength $K=-0.01\omega$ (with $\omega=\pi$). The black solid line represents the exact full quantum simulation (Hilbert space truncation $N=200$), while the red dashed line shows the TWA prediction. Note that due to the broad initial phase-space distribution, the anharmonic shearing causes the expectation value $\expval{e^{2i\eta\hat{q}}}$ to oscillate through zero and become negative. The modulus operation rectifies these sign reversals, resulting in the observed sharp dips and non-sinusoidal rebounds. The bottom panel displays the residual difference $\Delta V = V_{\text{quant}} - V_{\text{TWA}}$, which remains bounded within $\pm 0.03$.
    \textbf{(b)} Snapshots of the Wigner function $W(q,p)$ from the full quantum simulation at times $t=0, 0.9, 1.5, 3.0$. Note the emergence of negative quasi-probability regions (blue areas) due to quantum interference.
    \textbf{(c)} Corresponding phase-space distributions from TWA simulation. While TWA accurately captures the macroscopic shearing and geometric distortion, it yields a strictly non-negative distribution, failing to reproduce the microscopic quantum negativity observed in (b).}
    \label{fig:twa_validation}
\end{figure*}

To rigorously benchmark the validity of the TWA for our experimental parameters, we compared its predictions against exact quantum numerical simulations. We focused on the visibility observable $V(t)=\abs{\expval{e^{2i\eta\hat{q}}}}$. As illustrated in Fig.~\ref{fig:twa_validation}(a), we select a demanding parameter regime for validation: the initial state is a squeezed thermal state ($\bar{n}=1.0$, $S=-1.0$) subject to Kerr nonlinearity. Exact quantum dynamics were computed using a truncated Fock basis with dimension $N=200$. In this regime, the broad initial phase-space distribution enhances the sensitivity to anharmonic shearing, causing the expectation value $\expval{e^{2i\eta\hat{q}}}$ to cross zero and undergo sign reversals. Consequently, the visibility $V(t)$ rectifies these negative excursions, manifesting as sharp dips and non-sinusoidal rebounds. 

\begin{figure*}[t]
    \includegraphics[width=\textwidth]{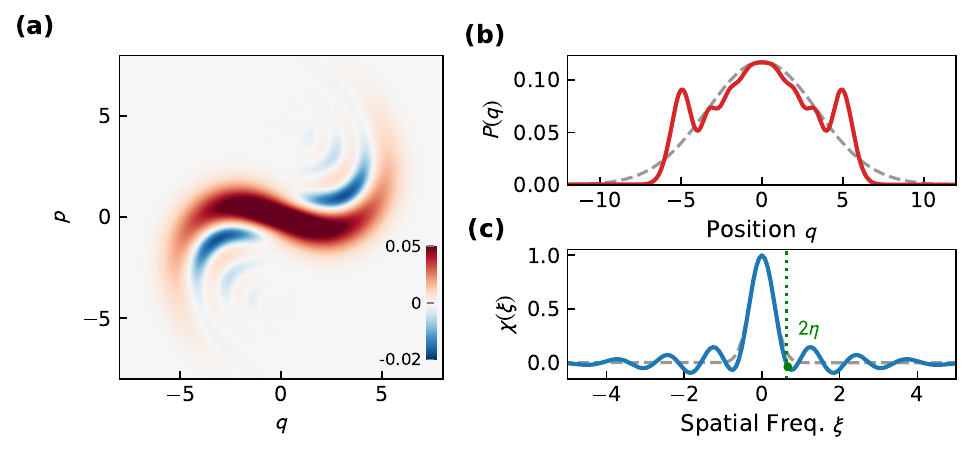}
    \caption{\textbf{Mechanism of visibility nodes arising from Kerr-induced non-Gaussianity.}
    \textbf{(a)} Wigner function $W(q,p)$ of the squeezed thermal state after evolution in the anharmonic potential. The simulation employs the best-fit parameters derived from the global Bayesian fit (see Fig.~4 of the main text and Sec.~\ref{sec:mcmc_fit}), assuming an initial thermal occupation $\bar{n}=0.5$. This state corresponds to the shallow-trap duration $\Delta T=22$~\si{\micro s} shown in Fig.~3(a) of the main text. The distribution exhibits characteristic phase-space shearing and negative quasi-probability regions (blue).
    \textbf{(b)} Corresponding spatial probability distribution $\mathcal{P}(q)$ deviates significantly from the Gaussian profile predicted by the harmonic approximation (gray dashed line).
    \textbf{(c)} Characteristic function $\chi(\xi)$, defined as the Fourier transform of $\mathcal{P}(q)$. The non-Gaussian distortion in real space induces oscillations in reciprocal space (blue solid line), contrasting with the strictly positive Gaussian reference (gray dashed line). The green dotted vertical line marks the interferometer's probe frequency $\xi=2\eta$. At this specific evolution time, the characteristic function takes a negative value $\chi(\xi)\approx-0.036$, implying that a zero-crossing point necessarily exists during the transition from the initial positive value to this negative amplitude.
    }
    \label{fig:characteristic}
\end{figure*}

These distinct zero-crossings, mirroring the features observed in Fig.~3(a,b) of the main text, arise fundamentally from the non-Gaussian distortion of the probability distribution $\mathcal{P}(q)$. Since the characteristic function $\chi(\xi)$ is the Fourier transform of $\mathcal{P}(q)$, the Kerr-induced distortion detailed in Fig.~\ref{fig:characteristic} forces $\chi(\xi)$ to oscillate and develop negative values. This behavior stands in contrast to the harmonic limit, where the Fourier transform of a Gaussian distribution remains strictly positive. Consequently, the visibility $V=\abs{\chi(2\eta)}$ vanishes specifically when the probe parameter $2\eta$ coincides with a zero node, while rectifying the negative excursions into the observed rebounds.

Despite these complex dynamics, the TWA prediction (red dashed line) exhibits excellent agreement with the full quantum result (black solid line). The residual deviation $\Delta V$ remains below 0.03 in the first 5 oscillation period. This validation justifies utilizing TWA as a computationally efficient forward model for the high-dimensional global Bayesian parameter estimation (Sec.~\ref{sec:mcmc_fit}). While TWA inherently yields a positive phase-space distribution [Fig.~\ref{fig:twa_validation}(c)], its accuracy in modeling the visibility observable ensures robust extraction of the system parameters ($S_1$, $K$). These extracted parameters are subsequently employed in the full quantum simulations presented in the main text [Fig.~4(a)] and Fig.~\ref{fig:twa_validation}(b) to reveal the genuine non-Gaussian resources and Wigner negativity.

%% file: sup_section/correction_sec1.tex
The final visibility $V(t)$ plotted in Fig.~3 of the main text results from a two-step data correction procedure designed to isolate the intrinsic phase-space dynamics from technical noise. While the interference visibility $V$ is conceptually defined through the normalized asymmetry of photon counts between the two output ports (Eq.~\eqref{eq:AsymmetryDef}), extracting $V$ by directly fitting the experimental asymmetry ratio is statistically suboptimal. The division of fluctuating photon counts distorts the underlying Gaussian noise statistics, resulting in a ratio distribution with heavy tails, particularly in the low-photon regime. Furthermore, post-interference imperfections, such as differential quantum efficiencies of the single-photon detectors, introduce offsets that complicate the simple asymmetry model.

Instead, we rigorously extract the raw visibility via a global joint fit to the raw photon counts, modeled as $N_\text{A}(\phi) = C_\text{A}[1+V\cos(\phi+\phi_0)]$ and $N_\text{B}(\phi) = C_\text{B}[1-V\cos(\phi+\phi_0)]$. The independent parameters $C_\text{A}$ and $C_\text{B}$ inherently decouple static detection imbalances from the interference contrast $V$. To resolve phase-amplitude degeneracy in low-visibility signals, we employ a sequential Bayesian strategy: the reference phase $\phi_\text{ref}$ derived from high-visibility calibration data serves as a Gaussian prior for the squeezed-state fit, ensuring rigorous uncertainty propagation into the final visibility $V_\text{meas,data}$.

Subsequently, we correct for pre-interference multiplicative suppression, arising primarily from photometric imbalance (e.g., unequal objective collection and single-mode fiber coupling efficiencies) and classical phase jitter, by normalizing the data visibility against interleaved ground-state calibration measurements. The resulting squeezed-thermal visibility, $V_\text{st}$, is defined as:
\begin{equation}
    V_\text{st} \equiv V_\text{meas,data} \times \frac{V_{\text{SQL},0}^{2\bar{n}_\text{calib}+1}}{V_\text{meas,calib}}
    \label{eq:V_st_definition}
\end{equation}
Here, the correction factor (the fraction) estimates the inverse of the technical contrast reduction. The numerator, $V_{\text{SQL},0}^{2\bar{n}_\text{calib}+1}$ (based on the thermal state variance derived in Eq.~\eqref{eq:V_th_theory}), represents the ideal physical visibility of the calibration state given its independently measured temperature $\bar{n}_\text{calib}$. This normalization isolates the visibility of the squeezed state from technical artifacts while explicitly preserving the thermal dependence ($\bar{n}_\text{data}$) of the target state. The quantities $V_\text{st}$ and $\bar{n}_\text{data}$ thus serve as the rigorous input for the physical modeling described in the next section.

%% file: sup_section/correction_sec2.tex
\label{sec:mcmc_fit}
To rigorously quantify the system parameters and decouple the intrinsic squeezing dynamics from thermal decoherence, we perform a global Bayesian joint fit across the two datasets  presented in Fig.~3. This analysis employs a unified physical model parameterized by the vector $\bm{\theta}=\{S_1, K/\omega_1,t_{\text{off},\Delta t},t_{\text{off},\Delta T},\omega_1\}$. The system dynamics are simulated using the TWA with an ensemble of $N=5\times10^4$ phase-space trajectories. The model incorporates the trap's intrinsic Kerr nonlinearity ab initio via the Hamiltonian $\hat{H}=\omega\hat{n}+K\hat{n}^2$, effectively capturing the non-Gaussian phase-space shearing and resultant decoherence. To ensure numerical stability and a smooth likelihood landscape for the sampler, we use a fixed ensemble of initial phase-space samples for all likelihood evaluations.

Each data point is associated with an independently measured initial thermal occupation $\bar{n}_i$ with uncertainty $\sigma_{\bar{n},i}$. To avoid the systematic bias, we explicitly marginalize over this thermal uncertainty. The likelihood function $\mathcal{L}(\bm{\theta})$ is constructed by comparing the raw experimental data $V_{\text{st},j}$ to the marginalized model prediction:
\begin{equation}
    \ln\mathcal{L}(\bm{\theta})\propto -\frac{1}{2}\sum_j\frac{[V_{\text{st},j}-\expval{\expval{V(t_j,\bm{\theta})}}_{\bar{n}}]}{\sigma_{V,j}^2}
\end{equation}
Here, the marginalized model prediction $\expval{\expval{V}}_{\bar{n}}$ is computed by averaging the TWA-simulated visibility over the experimental thermal error distribution:
\begin{equation}
    \expval{\expval{V(t_j,\bm{\theta})}}_{\bar{n}} = \int \dd \bar{n}\, P(\bar{n}\mid \bar{n}_j,\sigma_{\bar{n},j}) V(t_j,\bar{n},\bm{\theta})
\end{equation}
where $V(t_j,\bar{n},\bm{\theta})$ represents the visibility calculated from the TWA ensemble for a specific thermal occupation $\bar{n}$. In our numerical implementation, we efficiently approximate this integral by sampling the initial thermal condition $\bar{n}$ for each trajectory directly from the error distribution $P(\bar{n})$, thereby propagating the temperature uncertainty into the final observable without nested integration.

The parameter space is explored using Markov Chain Monte Carlo (MCMC) sampling \cite{foreman2013emcee}. We enforce a physical consistency constraint where the frequency ratio is determined by the squeeze parameter via the diabatic limit relation $\omega_2/\omega_1=\exp(-2S_1)$. The base trap frequency is constrained by a Gaussian prior centered at the experimentally measured value $\omega_1/2\pi=37.8(3)$~\si{kHz}, while the anharmonicity $K/\omega_1$ is restricted to non-positive values. The resulting posterior distributions yield a single-quench squeeze parameter of $S_1=0.50(1)$, a dimensionless Kerr coefficient of $K/\omega_1=-0.011(2)$, and a fitted base trap frequency of $\omega_1/2\pi=37.9(3)$~\si{kHz}.

To visualize the consistency between our global physical model and the experimental data, we plot the theoretical dynamics in Fig~3(a,b) using extracted optimal parameters $\theta_\text{best}$. The solid curves serve as a visual guide, representing the predicted evolution at a nominal thermal occupation of $\bar{n}=0.5$, while the shaded regions indicate the prediction bands covering the full range of experimentally measured occupations ($\bar{n}_\text{min}$ to $\bar{n}_\text{max}$) for the respective datasets.

\begin{figure*}[t!]
    \centering
    \includegraphics[width=\textwidth]{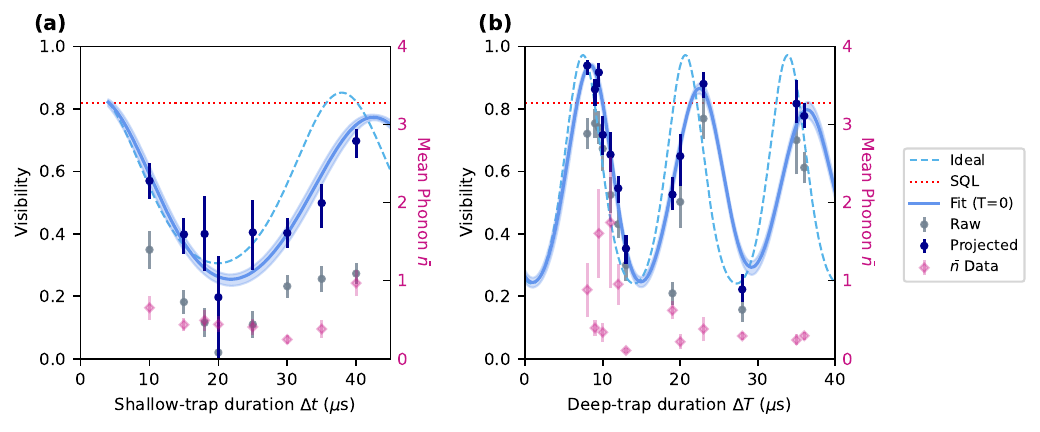}
    \caption{
    \textbf{Validation of intrinsic squeezing dynamics via thermal projection.}
    Comparison between the raw measured visibility (grey circles, $V_\text{st}$) and the thermally rescaled (dark blue circles, $V_\text{proj}$) projected to the zero-temperature limit ($\bar{n} = 0$). The solid blue curves represent the intrinsic quantum dynamics reconstructed from the global Bayesian fit, including trap anharmonicity. The light blue dashed curves show the ideal prediction for a harmonic potential ($K=0$), highlighting the deviation caused by the Kerr nonlinearity. Pink diamonds (plotted against the right axis) indicate the independent measurement of the mean thermal occupation $\bar{n}$ for each data points. The Standard Quantum Limit ($V_\text{SQL}$, red dotted line) is $0.819(1)$.
    \textbf{(a)} Dynamics in the shallow potential (squeezing generation) as a function of duration $\Delta T$.
    \textbf{(b)} Dynamics in the deep potential (evolution) as a function of duration $\Delta t$. 
    }
    \label{fig:projected}
\end{figure*}

We further validate the goodness of fit by verifying that the reconstructed intrinsic dynamics are consistent with the raw data after algebraically removing the thermal suppression. To achieve this, we calculate a generalized time-dependent scaling exponent $\gamma(t_i)=\ln V(t_i;0)/\ln V(t_i;\bar{n}_i)$ using the TWA model. This allows us to project each raw data point $V_\text{st}$ to the zero-temperature limit via $V_{\text{proj},i}=(V_{\text{st},i})^{\gamma(t_i)}$. As shown in Fig.~\ref{fig:projected}, these projected data points---with error bars representing the full propagation of measurement and scaling uncertainties---align closely with the ab initio reconstructed curve $V_\text{intrinsic}(t)$. This agreement confirms that our physical model accurately captures the scaling behavior, validating the parameters extracted from the global fit.

Subsequently, we perform an ab initio reconstruction of the system's dynamics to isolate the intrinsic quantum behavior from thermal fluctuations ($\bar{n}=0$). This simulation yields two key results based on the full uncertainty propagation. First, it generates the intrinsic visibility evolution presented in Fig.~4(b), where the solid curve corresponds to $\theta_\text{best}$ and the shaded region indicates the $1\sigma$ confidence interval derived from the full posterior ensemble. Second, to strictly quantify the generated non-classical resource, we examine the state immediately upon completion of the QEQ protocol. By simulating the dynamics under the full anharmonic Hamiltonian for each parameter set drawn from the posterior distribution, we diagonalized the resulting covariance matrix to extract the exact minimum quadrature variance. This rigorous approach, which accounts for the interplay between squeezing generation and trap anharmonicity, yields a total effective squeezing parameter of $S_\text{eff}=0.88(2)$, corresponding to a variance suppression of $-10\log_{10}(\sigma_\text{min}^2/\sigma_\text{vac}^2)=7.6(2)$~\si{dB}.

\begin{table*}[t!]
    \centering
    \renewcommand{\arraystretch}{1.5}
    \begin{tabular}{c | c | c }
        \hline\hline
        \textbf{Parameter} & \textbf{Description} & \textbf{Posterior (Median$\pm 1\sigma$}) \\
        \hline
        $S_1$ & Single-quench squeeze parameter (dimensionless) & $0.50(1)$ \\
        $K/\omega_1$ & Kerr coefficient (dimensionless) & $-0.011(2)$ \\
        $\omega_1/2\pi$ & Base trap frequency (\si{kHz}) & $37.9(3)$ \\
        $t_{\text{off},\Delta T}$ & Deep-trap duration offset (\si{\micro s}) & $4.2(6)$ \\
        $t_{\text{off},\Delta t}$ & Shallow-trap duration offset (\si{\micro s}) &  $0.5(3)$ \\
        \hline\hline
    \end{tabular}
    \caption{
        Posterior distributions for the 5-dimensional global Bayesian joint fit, based on the first-principles Kerr anharmonicity model. Uncertainties represent the 1-$\sigma$ confidence interval (derived from the 16th and 84th percentiles).
    }
    \label{tab:mcmc_posterior}
\end{table*}

\begin{figure*}[t!]
    \centering
    \includegraphics[width=\textwidth]{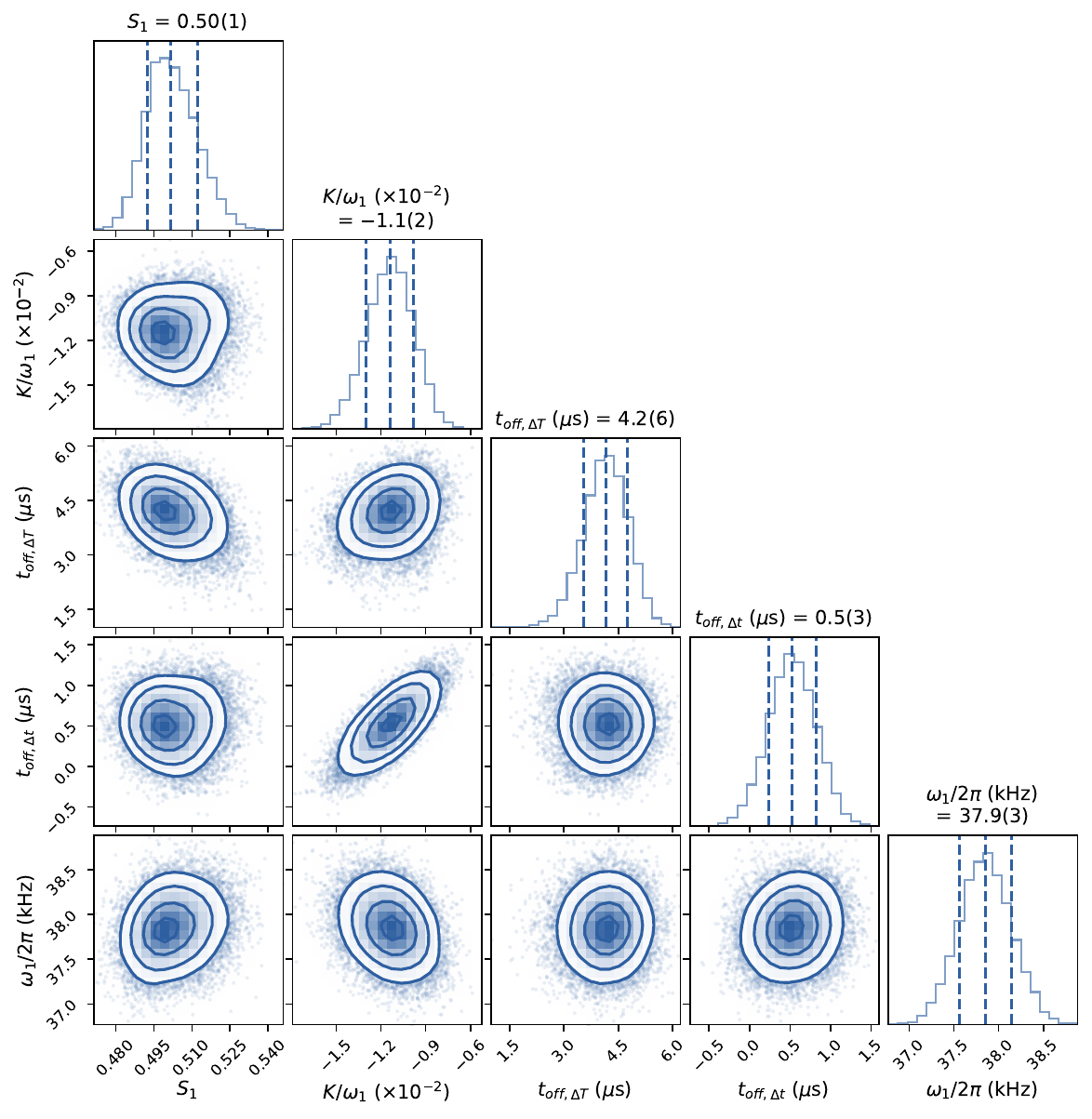}
    \caption{
    \textbf{Corner Plot of the 5D Global Bayesian Fit.}
    The diagonal panels show the 1D marginalized posterior distributions for each parameter: single-quench squeeze parameter $r$, dimensionless Kerr coefficient $K/\omega_1$, time offsets $t_{\text{off},\Delta t}$ and $t_{\text{off},\Delta T}$, and the base trap frequency $\omega_1$. The off-diagonal panels show the 2D joint posteriors, revealing correlations between parameter pairs. The median and 1-$\sigma$ confidence intervals (dashed lines) are shown on each 1D histogram, and the resulting statistics are summarized in Table~\ref{tab:mcmc_posterior}.
    }
    \label{fig:tomography}
\end{figure*}

%% file: sup_section/tomography.tex
\label{sec:tomography}
The main text concludes by noting that the interferometric method, having probed the characteristic function $\chi(k)$ at $k=(2\eta,0)$, can be extended to perform full Wigner function tomography of arbitrary motional states. This section details the protocol for such a reconstruction.

The protocol relies on the interferometric measurement's ability to extract the full complex value of the characteristic function, $\chi(k)=\tr(\rho e^{i(k_q\hat{q}+k_p\hat{p})})$. As derived in Sec.~\ref{sec:visibility_arbitrary}, a fit to the measured asymmetry signal $A(\phi_\text{int})$ yields both the visibility $V$ and the phase offset $\alpha$, thus determining the complex value $\chi=Ve^{i\alpha}$ at the specific probed $k$-vector.

 To sample the full $k$-space, the protocol combines passive phase-space rotation with the active quantum state engineering (the QEQ squeezing sequence) central to the main work. In the Heisenberg picture, this sequence transforms the fixed measurement operator $e^{2i\eta\hat{q}}$ into an operator that probes an arbitrary $k^\prime$-vector.
 
The generalized tomographic sequence commences with the preparation of the initial state $\rho$. This is followed by a free evolution $U(t_1)$, the total squeezing operation $\hat{S}(r)$, and a second free evolution $U(t_2)$. Finally, the interferometric measurement is performed on the resulting state by measuring the asymmetry signal $A(\phi_\text{int})$.

The complex amplitude $(V^\prime, \alpha^\prime)$ extracted from this measurement probes the initial state's characteristic function $\chi(k^\prime)$ at a transformed vector $k^\prime$. The scan of the second evolution time, $t_2$, samples $\chi(k)$ along an elliptical trajectory. The eccentricity of this ellipse is determined by the squeeze parameter $r$. The initial evolution time, $t_1$, determines the orientation of this sampled ellipse in $k$-space.

A systematic variation of $t_1$ and $t_2$, combined with the fixed $\hat{S}(r)$ operation, permits sampling of the 2D characteristic function within an annular $k$-space region, bounded by the semi-minor and semi-major axes of the sampling ellipse. The Wigner function $W(\hat{q}, \hat{p})$ is then recovered from this non-Cartesian dataset via an appropriate numerical method.

\begin{figure*}[p]
    \centering
    \includegraphics[height=0.5\textheight, keepaspectratio]{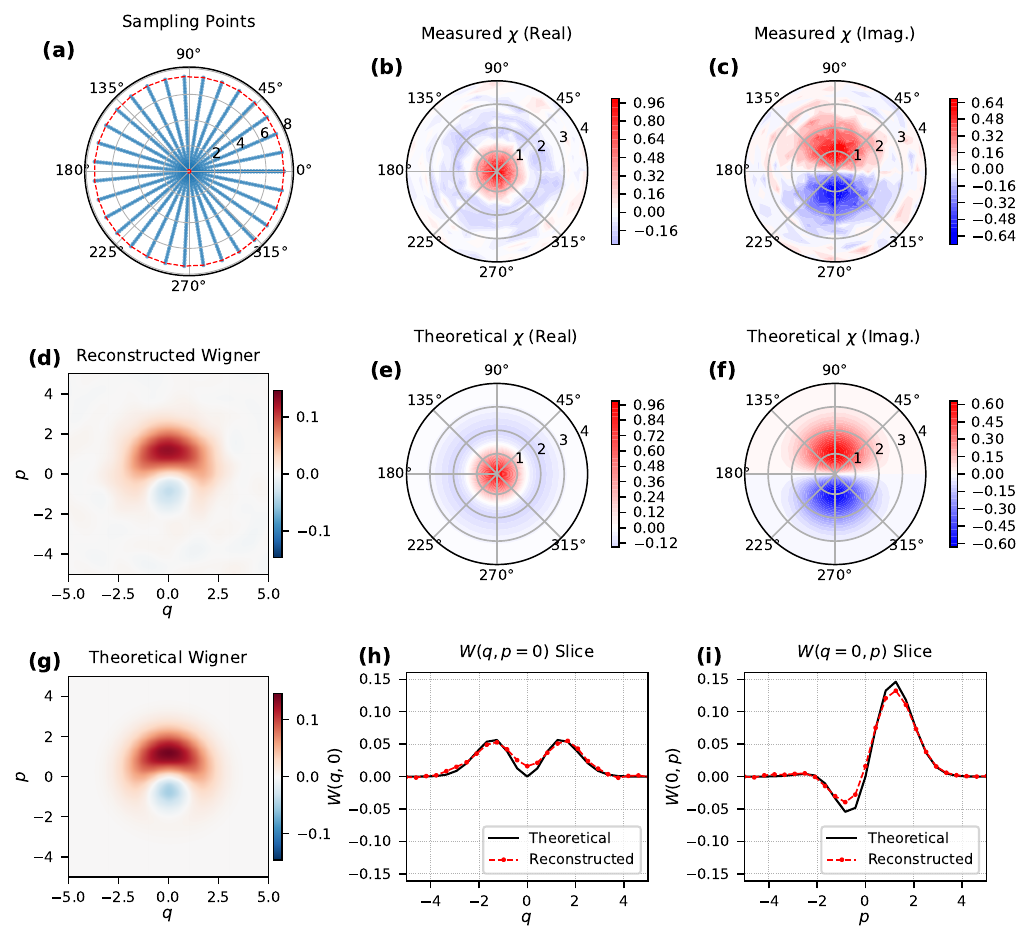}
    \caption{
    \textbf{Numerical Simulation of Interferometric Wigner Tomography.}
    Numerical simulation demonstrating the Wigner tomography protocol proposed in Sec.~\ref{sec:tomography} for a non-Gaussian target state $(\ket{0}+\ket{1})/\sqrt{2}$. The simulation assumes a base interferometer probe defined by a Lamb-Dicke parameter $\eta=0.5$ and a QEQ total squeeze parameter of $r=2$.
    \textbf{(a)} The non-Cartesian $k$-space sampling grid, corresponding to the annular region (spanning from 0.14 to 7.39) accessible via this QEQ protocol. \textbf{(b), (c)} Real and imaginary parts of the characteristic function $\chi(k)$ extracted from a simulated interferometric measurement. The deviations from the theoretical values (e, f) are caused by Poissonian projection noise introduced in the simulation. \textbf{(d)} The Wigner function $W(q,p)$ reconstructed from the noisy, non-Cartesian data in (b, c) via polar-to-Cartesian interpolation and a 2D inverse Fourier transform. \textbf{(e), (f), (g)} The corresponding ideal theoretical $\chi(k)$ and $W(q,p)$ for the target state. \textbf{(h), (i)} Slices of the reconstructed (dashed red) and theoretical (solid black) Wigner functions. The reconstruction robustly captures the state's qualitative structure. The quantitative deviations, such as the reduced sharpness and shallower negative features, are the expected consequence of the finite $k$-space sampling range, which acts as a low-pass filter on the $(q,p)$ space features.
    }
    \label{fig:tomography}
\end{figure*}

We numerically validate this protocol in Fig.~\ref{fig:tomography}. The simulation demonstrates that the reconstruction (Fig.~\ref{fig:tomography}(d)) successfully captures the principal qualitative, non-classical features of the target state  (Fig.~\ref{fig:tomography}(g)), notably its negative-valued region. The quantitative deviations, such as reduced sharpness and shallower negative features  (Fig.~\ref{fig:tomography}(h,i)), are the expected and fundamental consequence of the finite $k$-space sampling range (Fig.~\ref{fig:tomography}(a)). This limited Fourier-space coverage---determined by the interplay of the intrinsic Lamb-Dicke parameter $\eta$ and the applied squeeze parameter $r$---acts as a low-pass filter, which Poissonian projection noise contributes to the small-scale fluctuations.

This simulation confirms the protocol is a powerful, experimentally viable tool. In a practical implementation, the reconstruction fidelity could be significantly enhanced by further engineering. Increasing the $k$-space detection range (e.g., via larger $r$) would directly reduce the low-pass filtering, allowing for the recovery of finer state details. Moreover, this protocol relies on clean phase-space rotations to map the $k$-space. Its accuracy would therefore benefits from a trap with weaker Kerr anharmonicity than that characterized in Sec.~\ref{sec:kerr_anharmornicity}, thus minimizing the distortion of the sampling grid itself.

